\documentclass[aps,prb,showpacs,twocolumn]{revtex4}
\usepackage{amssymb}
\usepackage{amsmath,amsthm}
\usepackage{graphicx}
\usepackage{dcolumn}
\usepackage{bm}

\newtheorem{theorem}{Theorem}
\newtheorem{acknowledgement}[theorem]{Acknowledgement}

\input{tcilatex}

\begin{document}

\title{Fast quantum noise in Landau-Zener transition}
\author{V.L. Pokrovsky and D. Sun }
\affiliation{Department of Physics, Texas A\&M\ University, College Station, TX 77843}
\date{\today}

\begin{abstract}
We show by direct calculation starting from a microscopic model
that the two-state system with time-dependent energy levels in the
presence of fast quantum noise obeys the master equation. The
solution of master equation is found analytically and analyzed in
a broad range of parameters. The fast transverse noise affects the
transition probability during much longer time (the accumulation
time) than the longitudinal one. The action of the fast
longitudinal noise is restricted by the shorter Landau-Zener time,
the same as in the regular Landau-Zener process. The large ratio
of time scales allows solving the Landau-Zener problem with
longitudinal noise only, then solving the same problem with the
transverse noise only and matching the two solutions. The
correlation of the longitudinal and transverse noise renormalizes
the Landau-Zener transition matrix element and can strongly
enhance the survival probability, whereas the transverse noise
always reduces it. Both longitudinal and transverse noise reduce
the coherence. The decoherence time is inverse proportional to the
noise intensity. If the noise is fast, its intensity at which the
multi-quantum processes become essential corresponds to a deeply
adiabatic regime. We briefly discuss possible applications of the
general theory to the spin relaxation of molecular magnets.
\end{abstract}

\pacs{03.65.-w;33.80.Be;32.80.-t;}
\maketitle

\section{Introduction}

Landau-Zener (LZ) theory \cite{landau}, \cite{zener}, \cite{stueckelberg}
deals with a two-state system, whose levels vary in time by sweeping of some
physical parameter and would cross at some moment. Transitions between the
two states are induced by the transition matrix element $\Delta ,$ which in
static or adiabatic case causes the level repulsion (avoided crossing
according to the Wigner-Neumann theorem). The LZ theory is one of basic
dynamic problems of quantum mechanics. It has a broad range of applications
from the collision theory and quantum chemistry to disordered solids and
molecular magnets.

The LZ transitions in noisy environment recently attracted new attention in
connection with the problem of the qubit decoherence \cite{Legget}, \cite%
{weiss}. New experimental realization of qubits \cite{qubit-experiment} rose
again the question to what extent it is possible to minimize the decoherence
simultaneously maintaining sufficiently strong coupling to the external
signal.

The problem of the LZ transition in noisy environment has relatively long
history. One of the first considerations based on ideas of stochastic
trajectories belongs to Kasunoki \cite{kusunoki}. In the pioneering work
\cite{kayanuma85} Kayanuma calculated the transition amplitude in the
presence of a fast transverse Gaussian classical noise with a specific
(exponential) two-time correlation function\cite{footnote1}. This solution
was simplified and extended to general shape of correlation function by one
of the authors (VP) and N. Sinitsyn \cite{PS}.
%%CC00:by one of the authors?%%not changed 2/20.CC%%
In the same work a situation in which the transitions are produced by noise
as well as by regular Hamiltonian was considered. Pokrovsky and Scheidl \cite%
{PSch} calculated the two-time correlation function of the transition
probabilities for the LZ system subject to a fast classical transverse
noise. Longitudinal noise was considered by Kayanuma \cite{kayanuma84}, who
proved that strong fast longitudinal noise does not change the LZ transition
probability. Gefen \textit{et al.} \cite{gefen} and Ao and Rammer \cite%
{ao-rammer} considered more wide range of parameters and found the
situations in which the noise changes the transition probability. In the
work \cite{ao-rammer} a rather detailed analysis of different limiting cases
of temperature, coupling to the phonon bath, its spectral width and sweeping
rate was presented.
There occurred a controversy between the works \cite{gefen} and \cite%
{ao-rammer}. Generally, there is no complete agreement between different
authors on what happens in the adiabatic regime (very slow sweeping) in the
presence of the longitudinal noise. Motivated by this disagreement Kayanuma
and Nakamura performed a comprehensive analytical and numerical study of the
LZ transition in the presence of longitudinal noise \cite{kay-nak}. In
particular they obtained a formula for the case of strong decoherence which
is valid in both low-temperature and high-temperature limits. In all these
works the quantum nature of the longitudinal noise was taken in account.

Despite of significant progress a complete theory of the LZ transition in
noisy environment still does not exist. Theoretical works considered either
quantum longitudinal noise with transitions originated from the regular
transition matrix element or the classical transverse noise.
Quite recently Wubs \textit{et al}. \cite{wubs} have found a beautiful exact
formula for the transition probability of the 2-state system interacting
with the phonon bath at zero temperature\cite{footnote2}. The noise had both
longitudinal and transverse components. Their correlation and quantum nature
were substantial. No limitations to the noise strength and spectral width
were assumed. However, the limitation of zero temperature (phonon bath is in
the ground state) does not allow to extend these results to more realistic
situations.

The purpose of this article is to present a theoretical
description of the LZ system subject to a \textit{fast quantum}
noise which has both transverse and longitudinal components. It is
not yet complete theory, since it does not cover slow and
intermediate noise, but in its range of applicability it allows to
understand clearly all relevant physical regimes and phenomena. We
will show that, due to the fastness of the noise, the LZ
transition in the presence of the longitudinal noise and the
transitions due to the transverse noise are separated in time,
whereas the correlation between the transverse and longitudinal
noise leads to a renormalization of the regular transition matrix
element in the LZ Hamiltonian. For a moderately strong transverse
noise we derive master equations governing the population of the
two states and study their solution. If the transverse noise is
strong and also fast, the 2-state system falls into adiabatic
regime. The population of levels comes to the equilibrium with the
spin bath if the bath is in the state of thermal equilibrium. We
argue that a very strong noise is classical and adiabatic. In this
situation, as it was shown in \cite{PS}, the populations of the
two levels are equal.

The plan of the article is as follows. In section 2 we introduce
the Hamiltonian and characterize the noise. In section 3 we
present simple heuristic arguments resulting in master equations.
In section 4 we derive the master equations starting from
microscopic Hamiltonian for the case of the transverse noise only
and zero LZ transition matrix element. In section 5 we derive the
renormalization of the regular transition matrix element due to
correlation of longitudinal and transverse noise. In section 6 we
analyze the influence of the longitudinal noise. In section 7 find
the solution of master equation and study it. In section 8 we
match the solution of the master equation with the solution of the
LZ\ problem without transverse noise. Section 9 contains
discussion and conclusions. Here we compare our theory with that
by Wubs \emph{et al.} \cite{wubs}. We briefly analyze possible
applications of our theory to molecular magnets.

\section{Statement of the problem\label{statement}}

We consider a 2-state system interacting with a noisy environment. The
latter is a large system (bath) with a stationary density matrix. We neglect
the influence of the LZ transitions onto the state of the bath. For a
definiteness we will speak about the phonon bath, though it can include
other Boson excitations like spin waves, excitons, photons. Then the total
Hamiltonian of the system can be represented as follows:%
\begin{equation}
H=H_{2}+H_{b}+H_{int}  \label{total ham}
\end{equation}%
The term $H_{2}$ in equation (\ref{total ham}) represents the two-state
system:%
\begin{equation}
H_{2}=-\frac{\Omega \left( t\right) }{2}\sigma _{z}+\Delta \sigma _{x},
\label{2-state ham}
\end{equation}%
where $\sigma _{x}$ and $\sigma _{z}$ are Pauli matrices and $\Omega \left(
t\right) $ is the time-dependent frequency or the energy difference between
the so-called diabatic levels turning into zero at $t=0$; $\Delta $ is the
regular transition matrix element. If $t$ would be not time but some
parameter of the Hamiltonian, then non-zero $\Delta $ provides repulsion of
the adiabatic levels (the Wigner-Neumann theorem on avoided levels
crossing). For brevity we will call further the regular transition matrix
element $\Delta $ the LZ gap. Usually the linear approximation for the
frequency $\Omega \left( t\right) =\dot{\Omega}t$ proposed by Landau and
Zener is acceptable, but sometimes it is necessary to go beyond this
approximation. Namely, in real experiment the sweeping of $\Omega \left(
t\right) $ stops at some finite value, which can be not large in the
frequency scale of the problem. Therefore we will keep notation $\Omega
\left( t\right) $ throughout this article. The term $H_{b}$ in equation (\ref%
{total ham}) is the phonon bath Hamiltonian:%
\begin{equation}
H_{b}=\sum_{\mathbf{q}}\omega _{\mathbf{q}}b_{\mathbf{q}}^{\dag }b_{\mathbf{q%
}},  \label{bath ham}
\end{equation}%
where $b_{\mathbf{q}}$ and $b_{\mathbf{q}}^{\dag }$ are the operators of the
phonon annihilation and creation; $\omega _{\mathbf{q}}$ are the phonon
frequencies and $\mathbf{q}$ are their momenta. The interaction Hamiltonian
reads:%
\begin{equation}
H_{int}=u_{\Vert }\sigma _{z}+u_{\bot }\sigma _{x}  \label{int ham}
\end{equation}%
The Hermitian operators $u_{\Vert }$ and $u_{\bot }$ responsible for the
longitudinal and transverse noise depend linearly on the phonon operators.
Each of them is a sum of two Hermitian conjugated operators containing
either phonon annihilation or creation operators only:

\begin{equation}
u_{\alpha }=\eta _{\alpha }+\eta _{\alpha }^{\dag };~\eta _{\alpha }=\frac{1%
}{\sqrt{V}}\sum_{\mathbf{q}}g_{a}\left( \mathbf{q}\right) b_{\mathbf{q}%
};~\alpha =\Vert ,\bot  \label{noise operators}
\end{equation}%
where $g_{a}\left( \mathbf{q}\right) $ are complex coupling amplitudes and $%
V $ is the volume of the system supporting phonons. Quantum character of the
noise manifests itself in non-commutativity of operators $\eta _{\alpha }$
and $\eta _{\alpha }^{\dag }$. The problem consists in calculation of
transition and surviving probabilities for the 2-state system at a fixed
noise density matrix. In the absence of the noise, the transitions
amplitudes constitute the LZ transition matrix belonging to the SU(2) group
and depending on the dimensionless LZ parameter $\gamma _{LZ}=\Delta ^{2}/%
\dot{\Omega}$:

\begin{equation}
T_{LZ}=\left(
\begin{array}{cc}
\alpha & \beta \\
-\beta ^{\ast } & \alpha ^{\ast }%
\end{array}%
\right)  \label{LZ-t-matrix}
\end{equation}

\begin{equation*}
~\alpha =e^{-\pi \gamma _{LZ}}; ~\beta =-\frac{\sqrt{2\pi }\exp \left( -%
\frac{\pi \gamma _{LZ}}{2}+i\frac{\pi }{4}\right) }{\sqrt{\gamma _{LZ}}%
\Gamma \left( -i\gamma _{LZ}\right) }
\end{equation*}

If $\gamma _{LZ}$ is small, the system with the probability close to 1
remains in initial diabatic state; if $\gamma _{LZ}$ is large the system
with probability close to 1 proceeds along the adiabatic state, i.e. changes
the initial diabatic state to the alternative one. A characteristic time
necessary for the LZ transition is $\tau _{LZ}=\max \left( \Delta /\dot{%
\Omega},\dot{\Omega}^{-1/2}\right) $.

The influence of the noise onto the 2-state system is completely described
by the noise correlation functions:%
\begin{equation}
\left\langle \eta _{\alpha }\left( t\right) \eta _{\beta }^{\dag }\left(
t^{\prime }\right) \right\rangle =\frac{1}{V}\sum_{\mathbf{q}}g_{a}\left(
\mathbf{q}\right) g_{\beta }^{\ast }\left( \mathbf{q}\right) \left( n_{%
\mathbf{q}}+1\right) e^{i\omega _{\mathbf{q}}\left( t^{\prime }-t\right) }
\label{eta-etadag-t}
\end{equation}%
\begin{equation}
\left\langle \eta _{\alpha }^{\dag }\left( t\right) \eta _{\beta }\left(
t^{\prime }\right) \right\rangle =\frac{1}{V}\sum_{\mathbf{q}}g_{a}\left(
\mathbf{q}\right) g_{\beta }^{\ast }\left( \mathbf{q}\right) n_{\mathbf{q}%
}e^{i\omega _{\mathbf{q}}\left( t-t^{\prime }\right) }  \label{etadag-eta-t}
\end{equation}%
Here $n_{\mathbf{q}}=\left\langle b_{\mathbf{q}}^{\dag }b_{\mathbf{q}%
}\right\rangle $ are the average phonon occupation numbers and $\left\langle
...\right\rangle $ means averaging over the phonon bath ensemble, not
necessarily in thermal equilibrium. The Fourier-components of the
correlation functions (\ref{eta-etadag-t},\ref{etadag-eta-t}) read:%
\begin{equation}
\left\langle \eta _{\alpha }\eta _{\beta }^{\dag }\right\rangle _{\omega }=%
\frac{2\pi }{V}\sum_{\mathbf{q}}g_{a}\left( \mathbf{q}\right) g_{\beta
}^{\ast }\left( \mathbf{q}\right) \left( n_{\mathbf{q}}+1\right) \delta
\left( \omega -\omega _{\mathbf{q}}\right)  \label{eta-etadag-fourier}
\end{equation}%
\begin{equation}
\left\langle \eta _{\alpha }^{\dag }\eta _{\beta }\right\rangle _{\omega }=%
\frac{2\pi }{V}\sum_{\mathbf{q}}g_{a}\left( \mathbf{q}\right) g_{\beta
}^{\ast }\left( \mathbf{q}\right) n_{\mathbf{q}}\delta \left( \omega +\omega
_{\mathbf{q}}\right)  \label{etadag-eta-fourier}
\end{equation}%
Note that one of the two correlators contains only positive, whereas the
second one contains only negative frequencies. If the noise is in
equilibrium at temperature $T$, the Fourier-transforms of correlation
functions obey a simple relation $\left( \omega >0\right) $:%
\begin{equation}
\frac{\left\langle \eta _{\alpha }\eta _{\beta }^{\dag }\right\rangle
_{\omega }}{\left\langle \eta _{\alpha }^{\dag }\eta _{\beta }\right\rangle
_{-\omega }}=e^{\frac{\omega }{T}}  \label{equilibrium}
\end{equation}

Let us denote $\omega _{g}$ the range of frequencies in which the
coupling coefficients $g_{\alpha }\left( \mathbf{q}\right) $ do
not vanish. If the occupation numbers $n_{\mathbf{q}}$ are of the
same order of magnitude for all states within this region of
frequencies, then $\omega _{g}$ determines the spectral width of
the noise $\Delta \omega $. In some cases, for example at low
temperature $T\ll \omega _{g}$, there appears a second, smaller
scale of frequency ($T$). The noise correlation time is $\tau
_{n}=1/\Delta \omega $. It is different for the two correlation
functions $\left\langle \eta _{\alpha }\left( t\right) \eta
_{\beta }^{\dag }\left( t^{\prime }\right) \right\rangle $ and
$\left\langle \eta _{\alpha }^{\dag }\left( t\right) \eta _{\beta
}\left( t^{\prime }\right) \right\rangle $ at low temperature and
it is equal to $\omega _{g}^{-1}$ for both at high temperature. By
definition the noise is fast if $\tau _{n}\ll \min \left( \dot{\Omega}%
^{-1/2},\Delta ^{-1}\right) $ or $\Delta \omega \gg \max \left( \dot{\Omega}%
^{1/2},\Delta \right) $. Besides its spectral characteristics the noise is
characterized by its strength. The most natural measure of the noise
strength is the average square of its amplitude:%
\begin{equation}
\begin{split}
\left\langle u_{\alpha }^{2}\left( t\right) \right\rangle & =\left\langle
\eta _{\alpha }\left( t\right) \eta _{\alpha }^{\dag }\left( t\right)
\right\rangle +\left\langle \eta _{\alpha }^{\dag }\left( t\right) \eta
_{\alpha }\left( t\right) \right\rangle  \\
& =\frac{1}{V}\sum_{\mathbf{q}}\left\vert g_{a}\left( \mathbf{q}\right)
\right\vert ^{2}\left( 2n_{\mathbf{q}}+1\right)
\end{split}%
\end{equation}%
The noise is weak if $\left\langle u_{\alpha }^{2}\left( t\right)
\right\rangle \ll \dot{\Omega}$. Weak noise can be accounted as a
small perturbation to the LZ result. We call the noise moderately
strong if it obeys the inequality: $\left\langle u_{\alpha
}^{2}\left( t\right) \right\rangle \ll \tau _{n}^{-2}$. Though for
moderately strong noise the perturbation theory is generally
invalid at accumulation time scale determined in the next section,
we will show that it works during sufficiently small intervals of
time still longer than $\tau _{n}$. Most of our results relate to
the moderately strong noise. The noise is called strong if
$\left\langle u_{\alpha }^{2}\left( t\right) \right\rangle \geq
\tau _{n}^{-2}$. During another time scale, which we call the
relaxation time $\tau _{r\alpha }=\left( \left\langle u_{\alpha
}^{2}\left( t\right) \right\rangle \tau _{n}\right) ^{-1}$ (see \cite%
{kay-nak}, where it is called phase relaxation time)the
probability of finding the system in state 1 or 2 changes
significantly. When the noise is strong, the relaxation time
becomes less than the noise correlation time $\tau _{n}$. It is
convenient to introduce dimensionless coupling function $\lambda
_{\alpha \mathbf{q}}=\left\vert g_{a}\left( \mathbf{q}\right)
\right\vert /\left( \omega _{g}a^{3}\right) $,
where $a$ is the lattice constant. In condensed matter systems the values $%
\lambda _{\alpha \mathbf{q}}$ never become large. Migdal
\cite{migdal} argued that large coupling constants would lead to
the lattice instability and reconstruction. Though his arguments
related directly to the electron-phonon coupling, his idea is very
general. If $\lambda _{\alpha \mathbf{q}}$ are not large, a large
value of $\left\langle u_{\alpha }^{2}\left( t\right)
\right\rangle \tau _{n}^{2}$ can be reached only if phonon
occupation numbers $n_{\mathbf{q}}$ become large. In the
equilibrium case it means that the temperature must be large. Very
strong noise is classical irrespectively of its specific density
matrix. An analogue of the LZ parameter for the noise reads
\cite{PS}: $\gamma _{n\alpha }=\left\langle u_{\alpha }^{2}\left(
t\right) \right\rangle /\dot{\Omega}$. If it is small, the noise
brings only a small perturbation to the LZ picture; if it is
large, the occupation numbers of the 2-state system follow
adiabatically the instantaneous value of frequency. We will return
to this point later.

\section{Heuristic approach}

We start with an auxiliary problem specifying $\Delta =0$ and $\eta _{\Vert
}=0$, so that all transitions are only due to the transverse noise. Since $%
\tau _{n}\ll \dot{\Omega}^{-1/2}$, the instantaneous frequency of the
2-state system does not change during correlation time of the noise and it
can be considered as a constant. \ Thus, it is possible to calculate the
instantaneous rate of transition probability using the standard quantum
mechanical technique for transitions between stationary energy levels.

\begin{figure}[htpb]
\centering
\includegraphics[width=0.9\linewidth]{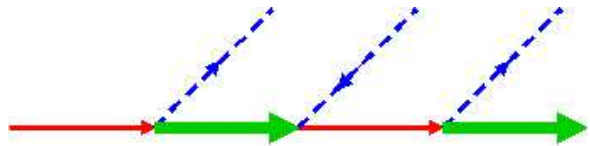}
\caption{Feynman graph for a 3-phonon process. Thin solid lines
correspond to the state 1; thick solid lines correspond to the
state 2; dashed blue lines correspond to phonons.}
\label{fig:fig1}
\end{figure}

\noindent For moderately strong noise the transition probability can be
calculated in the first Born approximation. Indeed, according to the Fermi
golden rule the rate of transition probability per unit time before the
level crossing reads:%
\begin{equation}
p_{1\rightarrow 2}\left( t\right) =2\pi \left\langle \eta _{\bot }^{\dag
}\eta _{\bot }\right\rangle _{\Omega \left( t\right) }  \label{golden}
\end{equation}%
In the framework \ of the considered model the next after
single-phonon is three-phonon transition shown in Fig. 1. Its
contribution to the transition probability reads:%
\begin{equation}
\begin{split}
& p_{1\rightarrow 2}^{\left( 3\right) }\left( t\right) =2\pi \\
& \times \left\langle \int \frac{d\omega _{1}d\omega _{2}}{\left( 2\pi
\right) ^{2}}\frac{\eta _{\omega _{1}}^{\dag }\eta _{\omega _{2}}^{\dag
}\eta _{\Omega (t)-\omega _{1}-\omega _{2}}^{\dag }}{\omega _{1}\omega _{2}}%
\times \mathrm{herm.~conj.}\right\rangle
\end{split}
\label{3-phono}
\end{equation}%
The 3-phonon contribution can be neglected if the noise is moderately
strong. In the same approximation it is possible to neglect the correction
to the transition frequency due to the interaction with the phonon bath.
Thus, the occupation numbers of the diabatic states $N_{1,2}$ at negative
time obey following master equation:%
\begin{equation}
\dot{N}_{1}=2\pi \left( -\left\langle \eta _{\bot }^{\dag }\eta _{\bot
}\right\rangle _{\left\vert -\Omega \left( t\right) \right\vert
}N_{1}+\left\langle \eta _{\bot }\eta _{\bot }^{\dag }\right\rangle
_{\left\vert \Omega \left( t\right) \right\vert }N_{2}\right)
\label{master-neg}
\end{equation}%
which must be complemented by conservation law $N_{1}+N_{2}=1$. For positive
time equation (\ref{master-neg}) must be modified as follows:%
\begin{equation}
\dot{N}_{1}=2\pi \left( -\left\langle \eta _{\bot }\eta _{\bot }^{\dag
}\right\rangle _{\left\vert \Omega \left( t\right) \right\vert
}N_{1}+\left\langle \eta _{\bot }^{\dag }\eta _{\bot }\right\rangle
_{\left\vert -\Omega \left( t\right) \right\vert }N_{2}\right)
\label{master-pos}
\end{equation}%
The noise produces transitions as long as its spectral width
exceeds the instantaneous frequency $|\Omega \left( t\right)| $.
The accumulation time estimated from this requirement is $\tau _{acc}=\left( \dot{\Omega}%
t_{n}\right) ^{-1}$ \cite{PS}. Since the noise is assumed to be fast the
accumulation time $\tau _{acc}$ is much longer than the noise correlation
time $\tau _{n}$. In real experiment the sweeping of frequency may stop or
saturate before the accumulation time is reached. The master equations
enable one calculating the occupation numbers at any time rather than
asymptotically at $t\rightarrow \infty $. \ The accumulation time is also
much longer than the LZ time $\tau _{LZ}$. Therefore, it is possible to
neglect the action of the noise during the LZ time interval $\left( -\tau
_{LZ}\text{,}\tau _{LZ}\right) $ and neglect the LZ gap $\Delta $ beyond
this time interval. It means that the action of the fast transverse noise
and of the LZ gap are separated in time as it was earlier shown for
classical noise \cite{PS}. The solution of the LZ problem without transverse
noise and the noise transition problem with zero LZ gap $\Delta $ should be
matched at some intermediate time. It will be done in Section \ref{combined}.

The action of the fast longitudinal noise is very different from that of the
transverse one. The longitudinal noise does not produce transitions in the
absence of the LZ gap. Therefore, its action is effectively limited to the
LZ time interval. In the next section we demonstrate that the fast
longitudinal noise must be sufficiently strong to produce a substantial
change in the LZ transition probability. Namely it must satisfy an
inequality $\left\langle u_{\Vert }^{2}\right\rangle \gtrsim \dot{\Omega}%
/\left( \Delta \tau _{n}\right) \gg \dot{\Omega}$. An analogous criterion
for the transverse noise is much more liberal: $\left\langle u_{\bot
}^{2}\right\rangle \gtrsim \dot{\Omega}$. For a comprehensive analysis of
the longitudinal noise action we refer the reader to the cited articles \cite%
{kay-nak}, \cite{ao-rammer}, \cite{gefen}. Beyond the LZ time interval the
classical longitudinal noise modulates the transverse noise by a factor $%
\exp \left( -i\int_{t_{0}}^{t}u_{\Vert }d\tau \right) $. Correlation
functions $\left\langle \eta _{\bot }\left( t\right) \eta _{\bot }^{\dag
}\left( t^{\prime }\right) \right\rangle $ must be substituted by $%
\left\langle \eta _{\bot }\left( t\right) \eta _{\bot }^{\dag }\left(
t^{\prime }\right) \exp \left( -i\int_{t^{\prime }}^{t}u_{\Vert }d\tau
\right) \right\rangle $. Neglecting the correlation between longitudinal and
transverse noise and employing the Gaussian statistics of the noise, one can
express the latter correlator as follows:

\begin{equation}
\begin{split}
&\left\langle \eta _{\bot }\left( t\right) \eta _{\bot }^{\dag
}\left( t^{\prime }\right) \exp \left( -i\int_{t^{\prime
}}^{t}u_{\Vert }d\tau \right)  \right\rangle =\left\langle \eta
_{\bot }\left( t\right) \eta _{\bot
}^{\dag }\left( t^{\prime }\right) \right\rangle \\
&\quad\times\exp \left[ -\frac{1}{2}\dint\limits_{t^{\prime
}}^{t}\dint\limits_{t^{\prime }}^{t}dt_{1}dt_{2}\left\langle u_{\Vert
}\left( t_{1}\right) u_{\Vert }\left( t_{2}\right) \right\rangle \right]
\label{Debye-Waller}
\end{split}%
\end{equation}

The transverse noise correlator decays rapidly when the modulus of
difference $\left\vert t-t^{\prime }\right\vert $ exceeds $\tau _{n}$.
Therefore, the value in the exponent in the right-hand side of equation (\ref%
{Debye-Waller}) can be estimated as $\left\langle u_{\Vert
}^{2}\right\rangle \tau _{n}^{2}\ll 1$. This estimates shows that the
longitudinal noise can be neglected beyond the LZ time interval. In the next
section we consider this question in more details.

\section{Derivation of Master equations.}

\begin{figure}[htpb]
\centering
\includegraphics[width=0.9\linewidth]{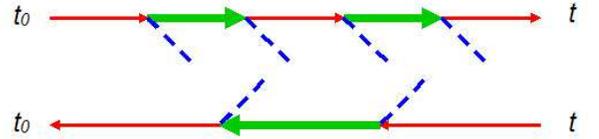}
\caption{An example of a term in the perturbation theory. Points correspond
to vertexes $V_{I}(t_{j})$.}
\label{fig:fig2}
\end{figure}
Our goal is to find the dependence of the occupation numbers $N_{\alpha
}~\left( \alpha =1,2\right) $ on time. The same problem can be formulated as
calculation of the average value of projectors $P_{\alpha }=\left\vert
\alpha \right\rangle \left\langle \alpha \right\vert =\frac{1}{2}\left( 1\pm
\sigma _{z}\right) $ \cite{kayanuma84}, \cite{ao-rammer}. We consider the
case $\Delta =0$, $u_{\Vert }=0$. The calculation will be performed in the
interaction representation with the diagonal time-dependent unperturbed
Hamiltonian $H_{0}=-\frac{\Omega \left( t\right) }{2}\sigma _{z}+H_{b}=$ $-%
\frac{\Omega \left( t\right) }{2}\left( \left\vert 1\right\rangle
\left\langle 1\right\vert -\left\vert 2\right\rangle \left\langle
2\right\vert \right) +H_{b}$ and the interaction Hamiltonian $V=u_{\bot
}\sigma _{x}=u_{\bot }\left( \left\vert 1\right\rangle \left\langle
2\right\vert +\left\vert 2\right\rangle \left\langle 1\right\vert \right) $.
Being transformed to the interaction representation, the interaction
Hamiltonian depends on time as follows:%
\begin{equation}
V_{I}\left( t\right) =u_{\bot }\left( t\right) \left( \left\vert
1\right\rangle \left\langle 2\right\vert e^{-i\int_{t_{0}}^{t}\Omega \left(
\tau \right) d\tau }+\left\vert 2\right\rangle \left\langle 1\right\vert
e^{i\int_{t_{0}}^{t}\Omega \left( \tau \right) d\tau }\right) ,
\label{intrepresent}
\end{equation}%
where%
\begin{equation}
u_{\bot }\left( t\right) =e^{iH_{b}\left( t-t_{0}\right) }u_{\bot }\left(
t_{0}\right) e^{-iH_{b}\left( t-t_{0}\right) }  \label{noise-intrepresent}
\end{equation}%
The time-dependent occupation numbers can be expressed in terms of the
evolution operator $U_{I}\left( t,t_{0}\right) $ in the interaction
representation:%
\begin{equation}
N_{\alpha }\left( t\right) =\mathrm{Tr}\left[ \rho _{0}U_{I}^{-1}\left(
t,t_{0}\right) P_{\alpha }U_{I}\left( t,t_{0}\right) \right]
\label{t-dependence}
\end{equation}%
where $\rho _{0}$ is the initial density matrix which is the direct product
of two independent density matrices $\rho _{0}=\rho _{2}\rho _{b}$, where
the first factor is the density matrix of the 2-state system and the second
once is the same for the bath. For calculation or these averages we employ a
simplified version of the Keldysh-Schwinger technique \cite{keldysh}, \cite%
{schwinger} used already for a similar purpose in \cite{kayanuma84}, \cite%
{ao-rammer}, \cite{kay-nak}. Each of the two evolution operators
is presented as a series of time ordered integrals. A general term
of such an expansion contains a product of two multiple time
integrals. With each time variable $t_{k}$ a vertex $V_{I}\left(
t_{k}\right) $ is associated. The product of vertices is ordered
chronologically in $U_{I}\left( t,t_{0}\right) $ and
antichronologically in $U_{I}^{-1}\left( t,t_{0}\right)
$. All operators of $V_{I}\left( t_{k}\right) $ belonging to $%
U_{I}^{-1}\left( t,t_{0}\right) $ are located on the left (``later'') than
corresponding operators belonging to $U_{I}\left( t,t_{0}\right) $. A
particular contribution is graphically depicted in Fig. 2 before the
averaging over phonon bath is performed. It consists of two lines both
starting at $t_{0}$ and ending at $t$. The upper line symbolizes $%
U_{I}\left( t,t_{0}\right) $ and the lower one symbolizes $U_{I}^{-1}\left(
t,t_{0}\right) $. Vertices on these lines correspond to the operators $%
V_{I}\left( t_{k}\right) $. Each vertex contains one phonon operator $%
u\left( t_{k}\right) $ and changes the state of the 2-state system. The
presence of the projection operator $P_{\alpha }$ in equation (\ref%
{t-dependence}) implies that the state closest to the final time $t$ on both
lines must be $\left\vert \alpha \right\rangle $. Applying the Wick's rule
for phonons, one should form all possible pairing of (phonon) noise lines%
\cite{footnote3}. However, in the case of fast moderately strong noise the
main contribution to the occupation numbers comes from the graphs without
crossing or overlapping of the phonon lines. An example of a graph without
overlapping is shown in Fig. 3. In comparison with elementary graphs without
phonon line overlapping or crossing (Fig. 4a,b) the contributions of the
graphs containing overlapping or crossing (Fig. 4c,d,) have additional small
factors of the order of $\left\langle u_{\bot }^{2}\right\rangle \tau
_{n}^{2}$ and can be neglected if the transverse noise is moderately strong.
Indeed the time interval between the ends of each phonon lines is about $%
\tau _{n}$. \newline

\begin{figure}[htpb]
\centering
\includegraphics[width=0.9\linewidth]{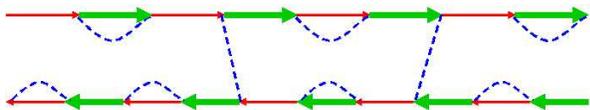}
\caption{A typical graph without phonon line crossings dominantly
contributing to the survival and transition probability.}
\label{fig:fig3}
\end{figure}

\begin{figure}[htpb]
\centering
\includegraphics[width=0.9\linewidth]{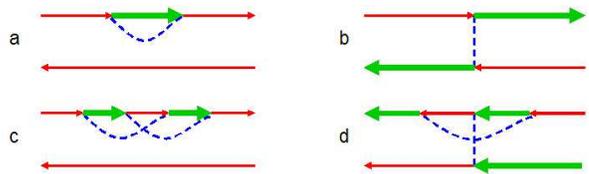}
\caption{Elementary graphs. a) b) without phonon line crossing; c) d) with
phonon line crossing.}
\label{fig:fig4}
\end{figure}

To derive the master equation we consider a set of graphs differing the
occupation numbers $N_{\alpha }\left( t+\Delta t\right) $ from those for $%
N_{\alpha }\left( t\right) $ with $\Delta t$ satisfying the following strong
inequality $\tau _{n}\ll \Delta t\ll \left( \left\langle u_{\bot
}^{2}\right\rangle \tau _{n}\right) ^{-1}$. First, there are graphs with
phonon lines connecting the interval $\left( t,t+\Delta t\right) $ with the
interval $\left( t_{0},t\right) $. Their contribution can be neglected since
it is relatively proportional to a small ratio $\tau _{n}/\Delta t$. The
contribution of $k$ non-overlapping or intersecting noise lines inside the
interval $\left( t,t+\Delta t\right) $ is proportional to $\left(
\left\langle u_{\bot }^{2}\right\rangle \tau _{n}\Delta t\right) ^{k}\ll 1$.
Therefore, the dominant contribution to the set comes from graphs containing
exactly zero or one line inside the interval $\left( t,t+\Delta t\right) $. Note
that it is a coarse-grain description: the master equation is invalid at
time scale $\tau _{n}$ and shorter. The graphical equation connecting $%
N_{\alpha }\left( t+\Delta t\right) $ with $N_{\alpha }\left( t\right) $ is
shown in Fig. 5.

\begin{figure}[htpb]
\centering
\includegraphics[width=0.9\linewidth]{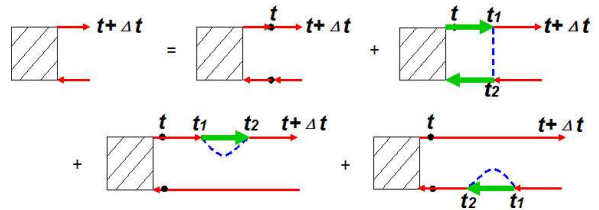}
\caption{Graphic equation connecting $N_{\protect\alpha }\left( t+\Delta
t\right) $ and $N_{\protect\alpha }\left( t\right) $ }
\label{fig:fig5}
\end{figure}

\noindent To find their analytical expression, we calculate the contribution
of the 3 elementary subgraphs for $\alpha =1$ shown in Fig. 6.

\begin{figure}[tbph]
\centering
\includegraphics[width=0.9\linewidth]{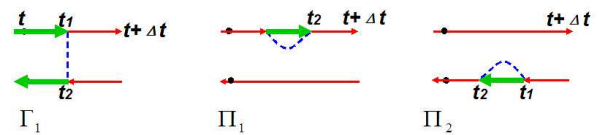}
\caption{Three elementary graphs with one phonon line in the
interval $t, t+\Delta t$.} \label{fig:fig6}
\end{figure}
They read:%
\begin{equation}
\begin{split}
\Gamma _{1}& =\dint\limits_{t}^{t+\Delta t}dt_{1}\dint\limits_{t}^{t+\Delta
t}dt_{2}\left\langle u_{\bot }\left( t_{1}\right) u_{\bot }\left(
t_{2}\right) \right\rangle e^{i\int_{t_{2}}^{t_{1}}\Omega \left( \tau
\right) d\tau } \\
& \approx 2\pi \left\langle u_{\bot }u_{\bot }\right\rangle _{\Omega \left(
t\right) }\Delta t
\label {up-down}
\end{split}%
\end{equation}%
\begin{eqnarray}%\begin{split}
\Pi _{1}& =-\dint\limits_{t}^{t+\Delta
t}dt_{1}\dint\limits_{t_{1}}^{t+\Delta t}dt_{2}\left\langle u_{\bot }\left(
t_{1}\right) u_{\bot }\left( t_{2}\right) \right\rangle
e^{-i\int_{t_{2}}^{t_{1}}\Omega \left( \tau \right) d\tau } \nonumber \\
& \approx -\Delta t\dint\limits_{-\infty }^{0}\left\langle u_{\bot }\left(
\tau \right) u_{\bot }\left( 0\right) \right\rangle e^{-i\Omega \left(
t\right) \tau }d\tau
\label{up-up}%\end{split}
\end{eqnarray}
\begin{eqnarray}%\begin{split}
\Pi _{2}& =-\dint\limits_{t}^{t+\Delta
t}dt_{1}\dint\limits_{t}^{t_{1}}dt_{2}\left\langle u_{\bot }\left(
t_{1}\right) u_{\bot }\left( t_{2}\right) \right\rangle
e^{-i\int_{t_{2}}^{t_{1}}\Omega \left( \tau \right) d\tau } \nonumber \\
& \approx -\Delta t\dint\limits_{0}^{\infty }\left\langle u_{\bot }\left(
\tau \right) u_{\bot }\left( 0\right) \right\rangle e^{-i\Omega \left(
t\right) \tau }d\tau
\label{down-down}%\end{split}
\end{eqnarray}

In this calculation we used the fastness of the noise $\left( \tau
_{n}\ll \Delta t,\text{ }\tau _{n}\ll \sqrt{\dot{\Omega}}\right) $
to substitute the integral in the exponent by $\Omega \left( t
\right) \left(
t_{1}-t_{2}\right) $ and to extend the integration over the difference $%
t_{1}-t_{2}$ to infinite limits. The graph 6a connects $N_{1}\left( t+\Delta
t\right) $ to $N_{2}\left( t\right) $, two others graphs connect $%
N_{1}\left( t+\Delta t\right) $ to $N_{1}\left( t\right) $.
Different signs in the contributions (\ref{up-down}) and
(\ref{up-up}, \ref{down-down}) are associated with the fact that
the vertex at the upper line contains a factor $-i$, whereas it
acquires the factor $+i$ at the lower line.

\begin{widetext}

Collecting all contributions together, we arrive at the Master
equations (\ref{master},  \ref{master-s-z}). 
We rewrite them%//master-neg & master-pos changed to master & master-s-z by dsun 3/19.
in a unified form:

\begin{equation}
\frac{dN_{1}}{dt}=2\pi
\left[\quad N_{2} \left[ \theta \left( \Omega \right)
\left\langle \eta \eta ^{\dag }\right\rangle _{\Omega }+\theta \left(
-\Omega \right) \left\langle \eta ^{\dag }\eta \right\rangle _{\Omega
}\right ]-N_{1} \left [ \theta \left( -\Omega \right) \left\langle \eta \eta
^{\dag }\right\rangle _{-\Omega }+\theta \left( \Omega \right) \left\langle
\eta ^{\dag }\eta \right\rangle _{-\Omega }\right ] \quad \right ]
 _{\Omega =\Omega\left( t\right) }
\label{master}
\end{equation}%
where $\theta \left( x\right) $ is the step function equal to 1 at positive
and 0 at negative arguments. The master equation looks simpler when the
variable $s_{z}=\frac{N_{1}-N_{2}}{2}=N_{1}-\frac{1}{2}$ is used:
\begin{equation}\begin{split}
\frac{ds_{z}}{dt}=2\pi \left[\quad -s_{z} \times\left[ \left\langle \eta \eta ^{\dag
}\right\rangle _{\left\vert \Omega \right\vert }+\left\langle \eta ^{\dag
}\eta \right\rangle _{-\left\vert \Omega \right\vert } \right]
-\frac{1}{2}
\mathrm{sign}\Omega  \times\left [ \left\langle \eta \eta ^{\dag }\right\rangle
_{\left\vert \Omega \right\vert }-\left\langle \eta ^{\dag }\eta
\right\rangle _{-\left\vert \Omega \right\vert } \right ] \quad\right] _{\Omega
=\Omega \left( t\right) }
\label{master-s-z}
\end{split}\end{equation}
\end{widetext}
The Fermi Golden Rule or the first Born approximation at a fixed
moment of time can be applied since the frequency variation during
the noise correlation is small and  the perturbation caused by the
noise in the corresponding stationary problem is weak. The
perturbation theory is valid if $\left\vert u_{\bot }\right\vert
\ll \left\vert \Omega \left( t\right) \right\vert $. Within the
accumulation time interval the instantaneous frequency is of the
same order of magnitude as the spectral width of the noise:
$\left\vert \Omega \left( t\right) \right\vert \sim \tau
_{n}^{-1}$. The same inequality ensures that the frequency exceeds
the width of the levels and the change of frequency due to the
interaction with the noise.

Now the question about the influence of the longitudinal noise onto the
master equation is in order. First we demonstrate that correlations of the
type $\left\langle u_{\Vert }\left( t\right) u_{\Vert }\left( t^{\prime
}\right) \right\rangle $ do not change the master equation. Indeed, let us
consider the influence of the longitudinal noise onto the difference between $%
N_{\alpha }\left( t+\Delta t\right) $ and $N_{\alpha }\left( t\right) $. In
analogy with the case of the transverse noise, the contribution of one%//comma added by dsun 3/19.
longitudinal phonon line inside the interval $\left( t,t+\Delta t\right) $
must be taken in account. This contribution does not depend on preceding
evolution of the system. Therefore, it is the same as it would be in the
absence of the transverse noise. Since the longitudinal noise does not
produce transitions in the absence of the transverse noise, the total
contribution of 3 graphs of Fig. 6 for longitudinal noise is zero.

\section{Renormalization of the LZ gap}

\begin{figure}[htpb]
\centering
\includegraphics[width=0.9\linewidth]{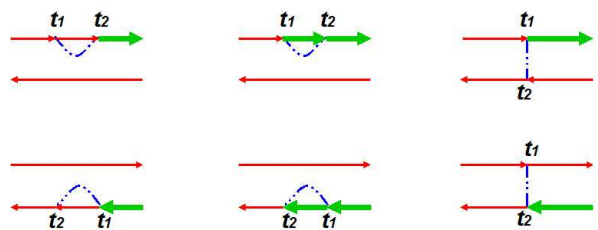}
\caption{Graphs containing mixed noise correlator (dash-dot line) and
responsible for the LZ gap renormalization.}
\label{fig:fig7}
\end{figure}

The problem of mixed correlations between transverse and longitudinal noise
is more subtle. The line of mixed correlation starts at one state, to say 1,
and ends at another one (2) as shown in Fig. 7. The self-energy part
associated with the graphs of the Fig. 7 reads:%
\begin{widetext}
\begin{equation}
\begin{split}
\Pi _{mix} =\left\vert 1\right\rangle \left\langle 2\right\vert
e^{-i\int_{t_{0}}^{t}\Omega \left( \tau \right) d\tau }
\times \left[
\dint\limits_{-\infty }^{0}\left\langle u_{\Vert }\left( t^{\prime }\right)
u_{\bot }\left( 0\right) \right\rangle dt^{\prime }-\dint\limits_{0}^{\infty
}\left\langle u_{\bot }\left( t^{\prime }\right) u_{\Vert }\left( 0\right)
\right\rangle dt^{\prime }\right]
+\mathrm{herm.~conj.}
\label{Mix-Pi}
\end{split}
\end{equation}
\end{widetext}This operator has the same form as the operator $\Delta \sigma
_{x}$ in the interaction representation. Thus, at a time scale much longer
than $\tau _{n}$, the mixed correlation renormalizes the LZ gap to a value%
\begin{equation}
\tilde{\Delta}=\Delta +i\dint\limits_{0}^{\infty }\left\langle \left[
u_{\bot }\left( t^{\prime }\right) ,u_{\Vert }\left( 0\right) \right]
\right\rangle dt^{\prime }  \label{Delta-renorm}
\end{equation}%
For the transformation of the integrals in equation (\ref{Mix-Pi}) into
integral in equation (\ref{Delta-renorm}) we employed the time-translation
invariance: $\left\langle u_{\Vert }\left( -t\right) u_{\bot }\left(
0\right) \right\rangle =\left\langle u_{\Vert }\left( 0\right) u_{\bot
}\left( t\right) \right\rangle $. Thus, the statement that one can neglect
the action of the transverse noise within the LZ time interval is not
completely correct: it is legitimate to neglect transverse-transverse
correlations, but the mixed correlations can significantly change the LZ gap
up to turning it into zero and changing its sign.

The commutator entering equation (\ref{Delta-renorm}) does not depend on the
phonon occupation numbers, i.e. on temperature. It is instructive to express
the renormalization of the LZ gap in terms of the phonon model (Section \ref%
{statement}, equations (\ref{noise operators}, \ref{eta-etadag-t}, \ref%
{etadag-eta-t})):%
\begin{equation}
\tilde{\Delta}-\Delta =-\frac{1}{V}\sum_{\mathbf{q}}\frac{g_{\Vert }\left(
\mathbf{q}\right) g_{\bot }\left( \mathbf{q}\right) }{\omega _{\mathbf{q}}}
\label{Delta-renorm-g}
\end{equation}

\section{Longitudinal noise}

Next we consider the action of the purely longitudinal noise
within the LZ time interval. For this problem the transverse noise
will be ignored. If the noise is classical, the proper diagonal
Hamiltonian is $H_{0}=\left( -\frac{ \Omega \left( t\right)}{2} +
u_{\Vert }\right) \sigma _{z}+H_{b}$, whereas the non-diagonal
part is $V=\Delta \sigma _{x}$. In the interaction representation
the non-diagonal part acquires the following form:
\begin{equation}
V_{I}\left( t\right) =\Delta \left( \left\vert 1\right\rangle \left\langle
2\right\vert e^{i\int_{t_{0}}^{t}\left( \Omega -2u_{\Vert }\right) d\tau }+%
\mathrm{herm.~conj.}\right)  \label{Delta-operator}
\end{equation}%
Calculation of the transition probability is very similar to that considered
above (see equation \ref{t-dependence} and Fig. 2), but the vertexes
correspond to $V_{I}\left( t\right) $ given by (\ref{Delta-operator}) (we
will call them $\Delta -$vertexes) and instead of connecting pairs of noise
amplitudes it is necessary to calculate average of a product $
%%CC04: pairs of noise amplitudes? CC%%
\dprod\limits_{j}\exp \left[ \pm 2i\int_{t_{0}}^{t_{j}}u_{\Vert
}\left( \tau _{j}\right) d\tau _{j}\right] $. Number of the signs
$-$ in the exponent is equal to the number of signs $+$.
Therefore, the dependence on the initial moment of time $t_{0}$
vanishes. The Gaussian statistics allows to calculate
the average of the product:%
\begin{widetext}
\begin{equation}\begin{split}
\left\langle \dprod\limits_{j}\exp \left[ \pm
2i\int_{t_{0}}^{t_{j}}u_{\Vert }\left( \tau _{j}\right) d\tau
_{j}\right] \right\rangle = \exp [
-2^{2k-1}\int_{t_0}^{t_{1}}d\tau _{1}...\int_{t_0}^{t_{2k}}d\tau
_{2k} \sum_{C}\left( \pm \left\langle u_{\Vert }(j_{1})u_{\Vert
}\left( j_{2}\right) \right\rangle \left\langle u_{\Vert
}(j_{3})u_{\Vert }\left( j_{4}\right) \right\rangle +...\right)]
\label{long-estimate}
\end{split}\end{equation}
\end{widetext}
where summation is performed over all possible divisions of
arguments $\tau _{j}$ into pairs. Each correlator vanishes if the
modulus of corresponding time difference $\left\vert \tau -\tau
^{\prime }\right\vert $ exceeds $\tau _{n}$. The substantial range
of integration over remaining variable $\left( \tau +\tau ^{\prime
}\right) /2$ is about $\tau _{LZ}=\Delta /\dot{\Omega}$.
Therefore the order of magnitude of the number obtained in the exponent (\ref%
{long-estimate}) after integration is $\sim \left( \left\langle u_{\Vert
}^{2}\right\rangle \tau _{n}\tau _{LZ}\right) ^{k}$. This value must be of
the order or larger than 1 to ensure a significant change of the transition
probability by the longitudinal noise. This requirement is equivalent to the
inequality $\left\langle u_{\Vert }^{2}\right\rangle \gtrsim \dot{\Omega}%
/\left( \Delta \tau _{n}\right) \gg \dot{\Omega}$. For the transverse noise
the analogues criterion is much softer \cite{PS}: $\left\langle u_{\bot
}^{2}\right\rangle \gtrsim \dot{\Omega}$. \

The quantum longitudinal noise does not commute with itself at different
moments of time. Therefore, it must be included into the non-diagonal
Hamiltonian. Then, besides of $\Delta -$vertexes one should consider the
noise vertexes. We obtain the same estimate considering the contribution of
an elementary graph with one noise line. However, for the fast longitudinal
noise it is possible to find not only an estimate, but equations for the
transition amplitudes. First we prove that the contribution of the
irreducible graphs\cite{footnote4}, which do not connect different branches
of the Keldysh contour, is zero. Indeed the remote ends of such graphs are
separated by the time interval of the order of $\tau _{n}\ll \tau _{LZ}$.
Therefore it is possible to integrate over the time difference at a fixed
"center of time" or "slow time" as we did in the case of the transverse
noise. In contrast to the latter case, for longitudinal noise the integral
does not depend on slow time since the longitudinal noise vertex in the
interaction representation does not contain time-dependent phase factor (it
connects identical states of the 2-state system). By the same reason it is
the same for states 1 and 2. Therefore in a slow time scale much longer than
$\tau _{n}$ (but much less than $\tau_{LZ}$) the contribution of such graphs
is proportional to unit operator for the 2-state system and can be
completely ignored. The irreducible noise graphs connecting different
branches of Keldysh contour form a 4-pole vertex (see Fig. 8), which by the
same reason does not depend on slow time and connects identical states at
each branch of the Keldysh contour. We denote this vertex $\Gamma $.

\begin{figure}[htpb]
\centering
\includegraphics[width=0.9\linewidth]{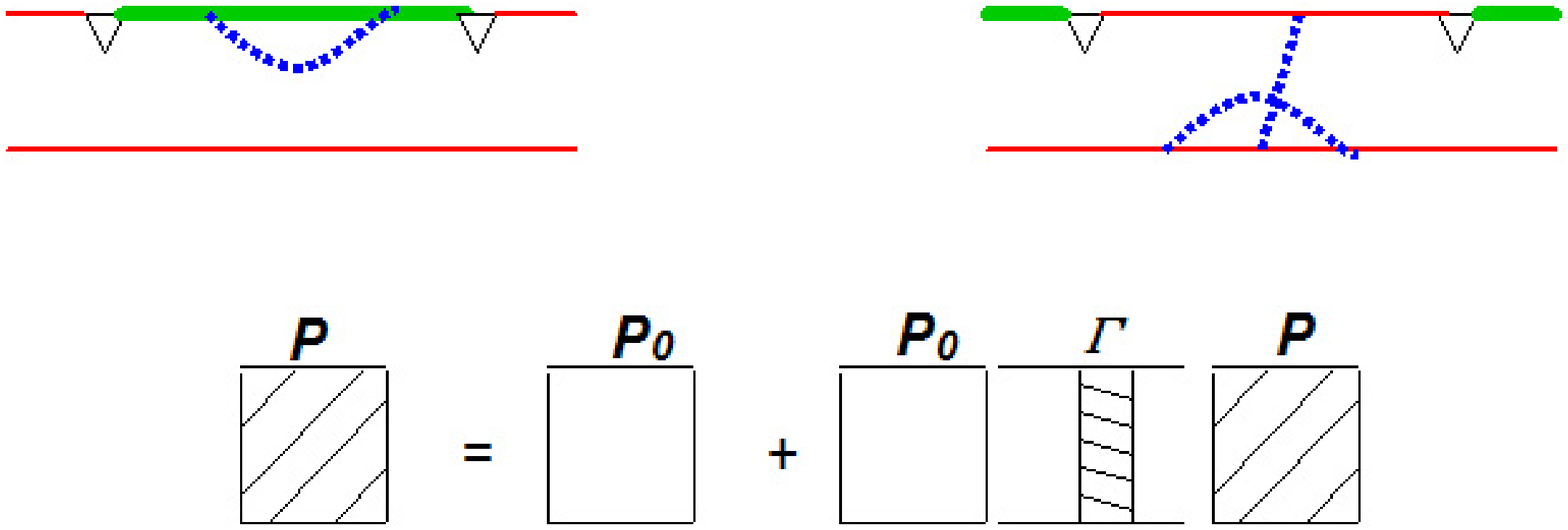}
\caption{Graphs containing the longitudinal noise only. Triangles correspond
to the LZ gap $\Delta$.\newline
a), graphs with one noise loop. In the slow time scale they are equivalent
to addition of a constant energy.\newline
b), graphs with a line connecting Keldysh branches. \newline
c), general graphic equation for $P$. See explanation in the text.}
\label{fig:fig8}
\end{figure}

Since the number of $\Delta -$vertexes between vertexes $\Gamma $ is
arbitrary we need to extend the number of amplitudes in consideration.
Namely, we define transition amplitudes $P_{\alpha \beta ,\alpha _{0}\beta
_{0}}\left( t,t_{0}\right) $ as the average value of the operator $%
\left\vert \alpha \right\rangle \left\langle \beta \right\vert $ at the
moment $t$ if the density matrix for 2-state system at the moment $t_{0}$
was $\left\vert \alpha _{0}\right\rangle \left\langle \beta _{0}\right\vert $%
. In the accepted approximation these amplitudes obey a system of linear
integral equations:

\begin{equation}
\begin{split}
&P_{\alpha \beta ,\alpha _{0}\beta _{0}}\left( t,t_{0}\right) = P_{\alpha
\beta ,\alpha _{0}\beta _{0}}^{\left( 0\right) }\left( t,t_{0}\right) \\
&\quad\quad -\Gamma \dint\limits_{t_{0}}^{t}P_{\alpha \beta ,\alpha ^{\prime
}\beta ^{\prime }}\left( t,t^{\prime }\right) P_{\alpha ^{\prime }\beta
^{\prime },\alpha _{0}\beta _{0}}^{\left( 0\right) }\left( t^{\prime
},t_{0}\right) dt^{\prime }  \label{long-equation}
\end{split}%
\end{equation}%
Here $P_{\alpha \beta ,\alpha _{0}\beta _{0}}^{\left( 0\right) }\left(
t,t_{0}\right) $ denotes the transition amplitude in the absence \ of noise.
The validity of equations (\ref{long-equation}) is limited by moderately
strong noise. Otherwise the amplitudes $P_{\alpha \beta ,\alpha _{0}\beta
_{0}}\left( t,t_{0}\right) $ vary significantly at a time scale $\tau _{n}$.
In this case more complicated equations with a non-local in time kernel $%
\Gamma $ and amplitudes depending on 4 time arguments must be used.

%A formal solution of equation (\ref{long-equation}) can be found
%by introduction of the operator $Q$ with the matrix elements
%$Q_{\alpha\beta ,t; \alpha^{\prime}\beta^{\prime}, t^{\prime}} =
%\theta (t-t^{\prime} \theta (t^{\prime} - t_0) P_{\alpha\beta;
%\alpha ^{\prime }\beta^{\prime} }\left( t,t^{\prime }\right)$ and
%operator $Q^{(0)}$ associated in analogues way with the kernel
%$P^{(0)}_{\alpha}\alpha ^{\prime }\left(t,t^{\prime }\right) $:
%\begin{equation}
%Q = \left( 1+\Gamma Q^(0)\right)^{-1}Q^(0)
%\end{equation}
%The kernel $P^{0}$ can be expressed in terms of the LZ evolution
%matrix $U_{LZ}\left( t,t^{prime}\right)$ as follows:
%\begin{equation}
%P^{(0)}_{\alpha \beta ,\alpha ^{\prime }\beta ^{\prime }}\left(
%t,t^{\prime }\right) = \langle\alpha^{\prime } | U_{LZ}\left(
%t,t^{prime}\right)|\alpha\rangle \langle\beta | U_{LZ}\left(
%t,t^{prime}\right)|\beta^{\prime}\rangle
%\end{equation}
%The matrix elements of the evolution matrix can be expressed in
%terms of the parabolic cylinder functions.

\section{Solution of the Master equation and noise diagnostic\label{Solution}%
}

The master equation (\ref{master-s-z}) allows an explicit solution:%//for deleted by dsun 3/19.
\begin{equation}
\begin{split}
s_{z}\left( t\right) =&s_{z}\left( t_{0}\right)\exp \left(
-\int_{t_{0}}^{t}f\left( \tau \right) d\tau \right) \\
&+\int_{t_{0}}^{t}g\left( t^{\prime }\right) \exp \left( -\int_{t^{\prime
}}^{t}f\left( \tau \right) d\tau \right) dt^{\prime },  \label{s-z-solution}
\end{split}%
\end{equation}%
where $f\left( t\right) =2\pi \left( \left\langle \eta \eta ^{\dag
}\right\rangle _{\left\vert \Omega \left( t\right) \right\vert
}+\left\langle \eta ^{\dag }\eta \right\rangle _{-\left\vert \Omega \left(
t\right) \right\vert }\right) $ and $g\left( t\right) =-\pi \mathrm{sign}%
\Omega \left( t\right) \left( \left\langle \eta \eta ^{\dag }\right\rangle
_{\left\vert \Omega \left( t\right) \right\vert }-\left\langle \eta ^{\dag
}\eta \right\rangle _{-\left\vert \Omega \left( t\right) \right\vert
}\right) $. For classical noise $g\left( t\right) =0$, and equation (\ref%
{s-z-solution}) reproduces the result obtained in the reference (\cite{PS}).
%%CC05: how to make the reference print as large as the normal font? CC%%
At zero temperature $\left\langle \eta ^{\dag }\eta \right\rangle
_{-\left\vert \Omega \left( t\right) \right\vert }=0$ \ and $g\left(
t\right) =-\frac{1}{2}\mathrm{sign}\Omega \left( t\right) f\left( t\right) $%
. In these two cases the measurement of occupation numbers or $s_{z}\left(
t\right) $ gives direct information on spectral power of noise. In classical
case we find $4\pi \left\langle \eta \eta ^{\dag }\right\rangle _{\left\vert
\Omega \left( t\right) \right\vert }=-\frac{d}{dt}\ln \left\vert
s_{z}\right\vert $; in purely quantum case ($T=0$) \ the relationship is
slightly more complicated: $2\pi \left\langle \eta \eta ^{\dag
}\right\rangle _{\left\vert \Omega \left( t\right) \right\vert }=-\frac{d}{dt%
}\ln \left\vert s_{z}+\frac{\mathrm{sign}\Omega \left( t\right) }{2}%
\right\vert $. In general case it is possible to find both spectral
functions $f\left( t\right) $ and $g\left( t\right) $ by performing two
series of measurements with different initial states. Thus, the 2-state
system is an ideal noise analyzer.

Next we consider regimes of very fast and very slow (adiabatic) frequency
sweeping. In the regime of fast sweeping $\left\langle \eta \eta ^{\dag
}\right\rangle \ll \dot{\Omega}$ the perturbation theory for equation (\ref%
{s-z-solution}) is valid. Indeed the integral $\int_{t_{0}}^{t}f\left( \tau
\right) d\tau $ can be rewritten in terms of spectral power as follows:%
\begin{equation}
\int_{t_{0}}^{t}f\left( \tau \right) d\tau =2\pi \dot{\Omega}%
^{-1}\dint\limits_{\Omega \left( t_{0}\right) }^{\Omega \left( t\right)
}\left( \left\langle \eta \eta ^{\dag }\right\rangle _{\left\vert \omega
\right\vert }+\left\langle \eta ^{\dag }\eta \right\rangle _{-\left\vert
\omega \right\vert }\right) d\omega  \label{time-frequency}
\end{equation}%
The integral in the r.-h. side of equation (\ref{time-frequency}) reaches
its maximum value, equal to the average square fluctuation $\left\langle
\eta \eta ^{\dag }+\eta ^{\dag }\eta \right\rangle $\ at $\Omega \left(
t_{0}\right) =-\infty $ and $\Omega \left( t\right) =+\infty $. If the
condition of fast sweeping is satisfied, the exponent in equation (\ref%
{s-z-solution}) can be expanded into a series over small noise parameter $%
\gamma _{n}=\left\langle \eta \eta ^{\dag }+\eta ^{\dag }\eta \right\rangle /%
\dot{\Omega}$. The variation $\Delta s_{z}\left( t\right) =s_{z}\left(
t\right) -s_{z}\left( t_{0}\right) $ is small at any time. In the leading
approximation it reads:%
\begin{equation}
\begin{split}
\Delta s_{z} (t) =-2\pi \dot{\Omega}^{-1}\dint\limits_{\Omega \left(
t_{0}\right) }^{\Omega \left( t\right) } [ \left\langle \eta \eta ^{\dag
}\right\rangle _{\left\vert \omega \right\vert } ( s_{z}\left( t_{0}\right)+%
\frac{\mathrm{sign}\omega }{2} ) \\
+\left\langle \eta ^{\dag }\eta \right\rangle _{-\left\vert \omega
\right\vert } ( s_{z}\left( t_{0}\right) -\frac{\mathrm{sign}\omega }{2} ) ]
d\omega  \label{fast sweep}
\end{split}%
\end{equation}
In the opposite regime of slow (adiabatic) sweeping the noise parameter $%
\gamma _{n}$ is large. In this case the exponents in equation (\ref%
{s-z-solution}) vary very rapidly allowing asymptotic calculation of $s_{z}$%
. However, in adiabatic regime it is simpler to start directly with the
Master equation (\ref{master-s-z}). Neglecting the time derivative in it, we
find the adiabatic solution:%
\begin{equation}
s_{z}\left( t\right) =\frac{g\left( t\right) }{f\left( t\right) }=-\frac{%
\mathrm{sign}\Omega \left( t\right) }{2}\frac{\left\langle \eta \eta ^{\dag
}\right\rangle _{\left\vert \Omega \left( t\right) \right\vert
}-\left\langle \eta ^{\dag }\eta \right\rangle _{-\left\vert \Omega \left(
t\right) \right\vert }}{\left\langle \eta \eta ^{\dag }\right\rangle
_{\left\vert \Omega \left( t\right) \right\vert }+\left\langle \eta ^{\dag
}\eta \right\rangle _{-\left\vert \Omega \left( t\right) \right\vert }}
\label{adiabatic}
\end{equation}%
If the photon bath is in equilibrium with temperature $T$, equation (\ref%
{adiabatic}) implies $s_{z}\left( t\right) =-\tanh \frac{\Omega \left(
t\right) }{2T}$. As it could be expected, at slow sweeping the two-state
system adiabatically accepts the equilibrium population with the temperature
of the bath. This conclusion shows that in the case of the quantum noise one
must be more careful with the asymptotic behavior of the time-dependent
frequency than in genuine LZ problem or even in the analogues problem with
the classical noise. In the latter problems the linear approximation for $%
\Omega \left( t\right) =\dot{\Omega}t$ was satisfactory. However, this
approximation may be invalid for the quantum noise if the sweeping stops
before the frequency $\Omega \left( t\right) $ reaches the spectral width of
the noise . In the opposite case the value $s_{z}\left( t\right) $ saturates
after $t=\tau _{acc}$. In classical adiabatic case $s_{z}\left( t\right) $
becomes zero after a short time $\tau _{tr}=\left( \left\langle \left\vert
\eta _{\bot }\right\vert ^{2}\right\rangle \tau _{n}\right) ^{-1}$. A
similar time scale for the longitudinal noise was introduced by Kayanuma and
Nakamura \cite{kay-nak}.

At the edge of the adiabatic regime $\gamma _{n}\sim 1$ the fast noise is
still moderately strong, i.e. $\sqrt{\left\langle \eta \eta ^{\dag
}\right\rangle }\ll \tau _{n}^{-1}$. It means that, when the noise becomes
strong $\sqrt{\left\langle \eta \eta ^{\dag }\right\rangle }\gtrsim \tau
_{n}^{-1}$, the system is already in deeply adiabatic regime. If the phonon
bath is in equilibrium, the 2-state system also is in equilibrium with the
noise. This equilibrium state is established in a time-independent, but
strongly non-linear system. The interaction of the two-level system
characterized by the time-independent frequency $\Omega $ with the strong
noise renormalizes the frequency and creates a finite width for each level.
The situation becomes simpler in the limit of very strong noise $\sqrt{%
\left\langle \eta \eta ^{\dag }\right\rangle }\gg \tau _{n}^{-1}\gtrsim
\Omega $. In this case the initial energy difference $\Omega $ between
levels can be neglected. The two states become equivalent and their%//Thew changed to The by dsun 3/19.
occupation numbers are equal (1/2), i.e. $s_{z}=0$. The same result can be
obtained from the fact that, as we already argued, the very strong noise
must be classical. Equation (\ref{adiabatic}) can be considered as an
interpolation between weak and very strong noise. Therefore, it gives a
reasonable description of intermediate regime.

\section{Transitions in the presence of the LZ gap and noise\label{combined}}

As we demonstrated earlier, for the fast moderately strong noise, the
effective time of the \ LZ transition due to the regular LZ gap $\tau
_{LZ}=\Delta /\dot{\Omega}$ is much less than the accumulation time $\tau
_{acc}=\left( \dot{\Omega}\tau _{n}\right) ^{-1}$. Therefore it is possible
to ignore the transverse noise within the LZ time interval $\left\vert
t\right\vert \lesssim \tau _{LZ}$ and to ignore the LZ gap $\Delta $ beyond
this interval. In this section we match the LZ solution modified by
longitudinal noise inside the LZ interval with the solution of the problem
with the transverse noise and $\Delta =0$ (see Section \ref{Solution})
beyond this interval. For this purpose we choose a time scale $t_{1}$ such
that $\tau _{LZ}\ll t_{1}\ll \tau _{acc}$ and first consider the solution (%
\ref{s-z-solution}) of the noise problem with $\Delta =0$ at $t=-t_{1}$. For
simplification we accept $t_{0}=-\infty $. \ Since $t_{1}\ll \tau _{acc}$ it
can be replaced by 0 in the solution (\ref{s-z-solution}) with high
precision $\sim t_{1}/\tau _{acc}$. Thus, at the left edge of the interval $%
\left( -t_{1},t_{1}\right) $ we find:%
\begin{equation}
\begin{split}
s_{z}\left( -t_{1}\right) \simeq s_{z}^{\left( -\right) }=\exp \left(
-\int_{-\infty }^{0}f\left( \tau \right) d\tau \right) s_{z}\left( -\infty
\right) \\
+\int_{-\infty }^{0}g\left( t^{\prime }\right) \exp \left( -\int_{t^{\prime
}}^{0}f\left( \tau \right) d\tau \right) dt^{\prime }  \label{s-z-}
\end{split}%
\end{equation}%
This value can be treated as an initial condition $s_{z}^{\left( -\right) }$
at $t=-\infty $ for the LZ problem with the longitudinal noise. If the
solution of this problem is known, the value $s_{z}^{\left( +\right) }$ at $%
t=+\infty $ can be calculated. The information necessary to make this
calculation effective is the knowledge of two numbers if there is no
coherence in the initial system. The density matrix $\rho ^{\left( +\right)
} $ at $t=+\infty $ is obviously a linear function of the initial density
matrix $\rho ^{\left( -\right) }$. There exists a linear 4$\times $4 matrix $%
\Lambda $ performing this transformation:%
\begin{equation}
\rho _{\alpha \beta }^{\left( +\right) }=\Lambda _{\alpha \beta ,\mu \nu
}\rho _{\mu \nu }^{\left( -\right) }  \label{Lambda}
\end{equation}%
The requirement that $\mathrm{Tr}\rho ^{\left( +\right) }=1$ if $\mathrm{Tr}%
\rho ^{\left( -\right) }=1$ implies the following equation: $\Lambda _{\alpha
\alpha,\mu \nu }=\delta _{\mu \nu }$. If $\rho _{12}^{\left( -\right) }=\rho
_{21}^{\left( -\right) }=0$, the equation (\ref{Lambda}) results in
following relationship between $s_{z}^{\left( +\right) }$ and $s_{z}^{\left(
-\right) }$:%
\begin{equation}
s_{z}^{\left( +\right) }=\left( \Lambda _{1}+\Lambda _{2}\right)
s_{z}^{\left( -\right) }+\left( \Lambda _{1}-\Lambda _{2}\right) ,
\label{s+s-}
\end{equation}%
where we introduced abbreviations $\Lambda _{1}$ and $\Lambda _{2}$ for $%
\Lambda _{11,11}$ and $\Lambda _{22,22}$, respectively. If the longitudinal
noise is absent or sufficiently weak, the LZ values for $\Lambda _{\alpha }$
are:%
\begin{equation}
\Lambda _{1}=\Lambda _{2}=\exp \left( -2\pi \gamma _{LZ}\right) -\frac{1}{2}
\label{LZ-Lambda}
\end{equation}%
If $\left\langle u_{\Vert }^{2}\right\rangle \ll \dot{\Omega}/\left( \Delta
\tau _{n}\right) $, the longitudinal noise is weak enough to neglect the
longitudinal-longitudinal correlations. Still the correlation of the
longitudinal and transverse noise can significantly change the effective LZ
gap (see equation (\ref{Delta-renorm})) .

The value $s_{z}^{\left( +\right) }$ from equation (\ref{LZ-Lambda}) serves
in turn as initial condition at $t=+0$ for the Master equation. Its solution
(\ref{s-z-solution}) at $t=+\infty $ leads to the final result:%
\begin{widetext}
\begin{equation}\begin{split}
&s_{z}  (+\infty  ) = ( \Lambda _{1}+\Lambda
_{2} ) e^{-2\pi \gamma _{\bot }}s_{z} ( -\infty  )
+
(\Lambda _{1}-\Lambda _{2} ) e^{-\pi \gamma _{\bot }} \frac{\pi }{\dot{\Omega}}
\times\\
&\quad\quad \dint\limits_{0}^{\infty }( \left\langle \eta \eta ^{\dag
}\right\rangle _{\Omega }-\left\langle \eta ^{\dag }\eta
\right\rangle _{-\Omega }) \exp [ -\frac{2\pi }{\dot{
\Omega}}\dint\limits_{\Omega }^{\infty }( \left\langle \eta \eta
^{\dag }\right\rangle _{\omega }+\left\langle \eta ^{\dag }\eta
\right\rangle
_{-\omega })
 d\omega ] [ ( \Lambda _{1}+\Lambda _{2}) e^{-\frac{4\pi }{\dot{\Omega
}}\int_{0}^{\Omega }( \left\langle \eta \eta ^{\dag }\right\rangle
_{\omega }+\left\langle \eta ^{\dag }\eta \right\rangle _{-\omega })
d\omega }-1] d\Omega
\label{s-z-+inf}
\end{split}\end{equation}%
\end{widetext}
Here $\gamma _{\bot }=\left\langle u_{\bot }^{2}\right\rangle /\dot{\Omega}$%
. We remind that the occupation numbers are related to $s_{z}$ as $%
N_{1,2}=1/2\pm s_{z}$. Below we write the survival probability for the case
when the longitudinal-longitudinal correlations can be neglected $%
\left\langle u_{\Vert }^{2}\right\rangle /\dot{\Omega}\ll \left( \Delta \tau
_{n}\right) ^{-1}$. In this case $s_{z}\left( -\infty \right) =1/2$, the
values $\Lambda _{1,2}$ are determined by equation (\ref{LZ-Lambda}) and
from equation (\ref{s-z-+inf}) we find:%
\begin{widetext}
\begin{equation}\begin{split}
&P_{1\rightarrow 1}=\frac{1}{2}\left[ 1+e^{-2\pi \gamma _{\bot
}}\left( 2e^{-2\pi \gamma _{LZ}}-1\right) \right] +  \frac{\pi }{\dot{\Omega}}\times \\
&\dint\limits_{0}^{\infty } (\left\langle \eta \eta ^{\dag
}\right \rangle _{\Omega }-\left\langle \eta^{\dag }\eta
\right\rangle _{-\Omega } ) \exp  [ -\frac{2\pi }{\dot{
\Omega}}\dint\limits_{\Omega }^{\infty } ( \left\langle \eta \eta
^{\dag}\right\rangle _{\omega }+\left\langle
\eta ^{\dag }\eta \right\rangle_{-\omega } ) d\omega ][  ( 2e^{-2\gamma _{LZ}}-1 ) e^{-\frac{4\pi }{\dot{\Omega}}
\int_{0}^{\Omega } ( \left\langle \eta \eta ^{\dag }\right\rangle
_{\omega }+\left\langle \eta ^{\dag }\eta \right\rangle _{-\omega } )
d\omega }-1 ] d\Omega
\label{survival}
\end{split}\end{equation}
\end{widetext}
To take in account the correlation between the longitudinal and transverse
noise one should replace the LZ gap $\Delta $ in the expression for $\gamma
_{LZ}=\Delta ^{2}/\dot{\Omega}$ by the renormalized value  from equations (\ref{Delta-renorm},\ref{Delta-renorm-g}).

\section{Discussion and conclusions}

In the case of weak transverse noise or very fast sweeping $\gamma _{\bot
}\ll 1$ equations (\ref{s-z-+inf}, \ref{survival}) turn into the result (\ref%
{s+s-}) and the LZ survival probability, respectively. In the opposite case
of the strong transverse noise or slow sweeping $\gamma _{\bot }\gg 1$ the
occupation numbers accept their stationary values at fixed instantaneous
frequency independently on the value of LZ parameter $\gamma _{LZ}$. The
classical noise corresponds to large phonon occupation numbers $n_{\mathbf{q}%
}$. In this case the operators $\eta $ and $\eta ^{\dag }$ commute; all
terms containing commutators $\left\langle \eta \eta ^{\dag }\right\rangle
_{\Omega }-\left\langle \eta ^{\dag }\eta \right\rangle _{-\Omega }$ can be
neglected. Then theory reproduces the result for classical fast noise \cite%
{PS}. It is instructive to compare equation (\ref{survival}) with the exact
survival probability for $T=0$ obtained in the recent article by Wubs
\textit{et al.} \cite{wubs}. At zero temperature the average value $%
\left\langle \eta ^{\dag }\left( t\right) \eta \left( t^{\prime }\right)
\right\rangle $ as well as its Fourier transform turns into zero. This fact
allows to calculate the integrals in equation (\ref{survival}). More
physically visible way of obtaining the same result is to keep in mind that
there is no live phonon at $T=0$ and%//live phonons changed to live phonon by dsun 3/19. 
only spontaneous emission of phonons
is possible. Therefore, if initially only the lower state was populated, the
phonon cannot be emitted before the levels crossing. This consideration
immediately gives $s_{z}^{\left( -\right) }=s_{z}\left( -\infty \right) =%
\frac{1}{2}$ and $s_{z}^{\left( +\right) }=\frac{1}{2}\left( 2e^{-2\gamma
_{LZ}}-1\right) $. Employing general equation (\ref{s-z-solution}), we find
the value $s_{z}\left( +\infty \right) =\frac{1}{2}\left[ 2\exp \left( -2\pi
\left( \gamma _{LZ}+\gamma _{\bot }\right) \right) -1\right] $ and the
survival probability:%
\begin{widetext}
\begin{equation}\begin{split}
 P_{1\rightarrow 1}=\exp \left( -2\pi \left( \gamma _{LZ}+\gamma
_{\bot }\right) \right) =
\exp [ -\frac{2\pi }{\dot{\Omega}} ( ( \Delta
-\frac{1}{V} \sum_{\mathbf{q}}\frac{g_{\Vert }\left(
\mathbf{q}\right) g_{\bot }\left( \mathbf{q}\right) }{\omega
_{\mathbf{q}}}) ^{2}+\left\langle \eta _{\bot }\left(
0\right) \eta _{\bot }^{\dag }\left( 0\right) \right\rangle
 )  ]
\label{survival-zero}
\end{split}\end{equation}
\end{widetext}This result with precision of notations coincides with the
exact result by Wubs \textit{et al.} \cite{wubs}, equations (6-8), obtained
without any limitations to the strength of noise and ratios of
characteristic time scales. Surprisingly the multiphonon processes as well
as the longitudinal-longitudinal noise correlations do not contribute at all
to the survival and transition probabilities even at very high noise
intensity. At zero temperature such a high noise level can be reached only
by enhancement of the coupling amplitudes. Though large coupling amplitudes
are physically implausible, as a mathematical model they are absolutely
legitimate. The fact that these high-intensity processes do not play role in
the transitions supports our speculations about possible extension of the
master equation beyond their range of validity at least as a reasonable
interpolation procedure.

Our theory is relevant to molecular magnets, first of all because the
condition of the noise fastness is perfectly satisfied in the experiment.
Indeed, the highest magnetic field rate used in the experiments with Fe$_{8}$
and Mn$_{12}$ was 10$^{3}$ Gs/s \cite{wernsdorfer}, \cite{sarachik}. This
rate corresponds to $\dot{\Omega}=10^{10}$s$^{-2}$. The lowest temperature
used in the cited measurements was about 0.05K. The dimensionless ratio of
the value $\sqrt{\dot{\Omega}}$ to the smaller of the noise spectral widths
is $\hbar \sqrt{\dot{\Omega}}/T\sim 10^{-5}$. \ The estimates of the noise
intensity is not so simple. The transverse noise is very weak for
transitions with a large change of the spin projection, for example, from
+10 to -10, since the standard  coupling of the deformations to spin can
change the spin projection only by $\pm 1,\pm 2$. Therefore, the transverse
noise for such a transition appears only in the 10-th order of the\
perturbation caused by acoustic phonons. However, for transitions with the
small change of the spin projection the transverse noise is of the same
order of magnitude as longitudinal one. The latter can be estimated from the
known value of magnetic anisotropy energy $E_{a}\simeq 60K$ (for Fe$_{8}$)
and the statistical weight of the phonons with the frequency smaller than
anisotropy energy $\sim \left( E_{a}/E_{D}\right) ^{3}$, where $E_{D}$ is
the Debye energy. An additional factor $E_{a}/E_{D}$ comes from the square
of coupling function $\left\vert g_{\mathbf{q}}\right\vert ^{2}$, which is
proportional to $q$ at small wave vectors. Thus, $\sqrt{\left\langle
u_{\Vert }^{2}\right\rangle }\sim E_{a}\left( E_{a}/E_{D}\right) ^{4}\ll
E_{a}=\hbar \omega _{g}$. This inequality shows that the condition of
moderately strong noise is well satisfied, whereas the value $\left\langle
u_{\Vert }^{2}\right\rangle /\hbar ^{2}\dot{\Omega}$ is large. This fact
together with the experimental fact that the hysteresis curve significantly
narrows at $T\sim 2-3K$ shows that the noise level is overestimated by this
simple formula and more accurate theory is necessary. However, there is no
doubt, that the thermal noise becomes important at a temperature of few
Kelvin and that the noise is fast and quantum. Though the longitudinal noise
does not produce transitions between diabatic levels, it is effective for
transitions between adiabatic levels \cite{ao-rammer}.

In conclusion, we derived the master equations for moderately strong fast
quantum transverse noise. We obtained its solution at any moment of time. We
demonstrated that the action of the regular LZ transition matrix element and
the longitudinal-longitudinal noise correlation is limited by the LZ time
scale, whereas the action of the transverse noise is accumulated during much
longer accumulation time. We showed that the mixed longitudinal-transverse
noise correlation leads to the renormalization of the LZ gap $\Delta $. In
the limiting case of adiabatic transverse noise the 2-state system
adiabatically follows its stationary state at an instantaneous value of
frequency independently on the value of LZ parameter. The separation of time
scales allows to derive exact transition probability with the LZ gap,
longitudinal and transverse noise taken in account simultaneously. The
transition probabilities depend explicitly on the noise commutator
reflecting the quantum nature of the noise. It plays an important role
especially in the adiabatic regime. In the extreme quantum regime at zero
temperature our result coincides with exact result by Wubs \textit{et al }.
\cite{wubs}. We argued that the strong-noise effects such as multiphonon
processes and change of frequency appear only in the adiabatic regime for
the fast noise and do not change substantially the transition probability.

\begin{acknowledgement}
Our thanks due to Drs. N. Sinitsyn, Y. Gefen and A. Abanov for useful
discussions and B. Dobrescu for participation in the initial stage of this
work and a valuable remark. This work was supported by the DOE under the
grant DE-FG02-06ER46278 and partly by the NSF under the grant DMR-0321572.
VP\ acknowledges the hospitality and support of the Max Planck Institut f%
\"{u}r Komplexen Sytemen, Dresden, during the Workshop on Mesoscopics,
August-September 2006.
\end{acknowledgement}


\begin{thebibliography}{99}
\bibitem{footnote1} The noise is called transverse if it produces
transitions between the two states in the absence of regular transition
matrix element.
\bibitem{footnote3} The Wick pairing ensures that the numbers of the state
alternations on the upper and the lower lines are either both even or both
odd. Therefore, the occupation numbers depend only on diagonal matrix
elements of the initial density matrix.
\bibitem{footnote4} Following the quantum field tradition, we call a graph
irreducible if it can not be separated into too disconnected parts (which
are not connected by a noise line).
\end{thebibliography}
\end{document}